# ALGORITHMIC CRIMINAL LIABILITY IN GREENWASHING: COMPARING INDIA, UNITED STATES, AND EUROPEAN UNION

Sahibpreet Singh[*] & Dr. Manjit Singh[**]

## Abstract

*AI-powered greenwashing has emerged as an insidious challenge within corporate sustainability governance. It exacerbates the opacity of environmental disclosures. It subverts regulatory oversight. This study conducts a comparative legal analysis of criminal liability for AI-mediated greenwashing across India, the US, and the EU. It exposes doctrinal lacunae in attributing culpability when deceptive sustainability claims originate from algorithmic systems rather than human actors. Existing statutes exhibit anthropocentric biases by predicating liability on demonstrable human intent, rendering them ill-equipped to address algorithmic deception. The research identifies a critical gap in jurisprudential adaptation, as prevailing fraud and environmental statutes remain antiquated vis-à-vis AI-generated misrepresentation. Utilising a doctrinal legal methodology, this study systematically dissects judicial precedents, statutory instruments, and regulatory directives, yielding promising results regarding the potential expansion of corporate criminal liability doctrines. Preliminary findings underscore the viability of strict liability models, the recalibration of corporate governance frameworks to incorporate AI accountability, and the institutionalisation of algorithmic due diligence mandates under ESG compliance regimes. Comparative insights reveal jurisdictional disparities in corporate culpability paradigms, with the EU Corporate Sustainability Due Diligence Directive (CSDDD) offering a potential transnational model for regulatory harmonisation. This study contributes to the discourse on AI ethics and environmental jurisprudence by advocating for a hybrid liability framework that integrates algorithmic risk assessment with legal personhood constructs. The implications necessitate a doctrinal evolution that fortifies juridical architectures against AI-driven environmental deception. The findings advocate for an interdisciplinary approach to AI regulation, ensuring algorithmic opacity does not preclude liability enforcement in sustainability-related misrepresentation.*

[*] LLM, Department of Laws, Guru Nanak Dev University, Amritsar.
[**] Assistant Professor, Department of Laws, Guru Nanak Dev University, Amritsar.

# I
# Introduction

Greenwashing denotes the systematic dissemination of spurious assertions or deceptive representations concerning the purported ecological integrity of corporate commodities or operational modalities.[1] This duplicitous modus operandi is meticulously orchestrated to entice environmentally cognizant consumers. It concurrently functions as an apparatus for inveigling investors. The overarching objective remains the fortification of a hegemonic commercial ascendancy. The significance of greenwashing within environmental law is profound.[2] It vitiates meticulously architected regulatory scaffolds instituted to engender bona fide sustainability. It precipitates consumer disillusionment. It engenders inequitable market distortions; it culminates in ecological debasement. It perpetuates untenable industrial practices that masquerade under an eco-centric veneer. The nomenclature was initially enunciated by environmentalist Westerveld in 1986.[3] His critique castigated the hospitality sector's mendacious sustainability expositions. Over time, greenwashing has evolved from mere marketing tactics to complex fraudulent schemes impacting regulatory compliance, consumer trust, and financial markets.[4] Greenwashing engenders corporate accountability, consumer protection, and climatic amelioration. A plethora of legislative enactments is meticulously designed to proscribe the dissemination of spurious ecological assertions. Nevertheless, the efficacious enforcement of these regulations remains an onerous endeavour. The protean subterfuges employed by corporate entities incessantly obfuscate juridical intervention and regulatory oversight.[5] The proliferating assimilation of artificial intelligence into corporate environmental governance has exacerbated the convolutions intrinsic to greenwashing. AI-driven mechanisms are now operationalising the automation of sustainability disclosures. They orchestrate the analytical dissection of ESG (Environmental, Social, and Governance) metrics.[6] They exert a formidable influence over consumer perception. However, these algorithmic faculties simultaneously capacitate AI to aggrandise or contrive eco-centric bona fides, leading to a new phenomenon, "AI-driven greenwashing."[7] AI can enable greenwashing in the following ways:

a) **Fabricated Sustainability Reports**: Machine learning models possess the computational dexterity to assimilate voluminous datasets and synthesise corporate sustainability expositions. However, the veracity of such disclosures is contingent upon the integrity of the underlying data. Manipulative recalibration, selective omission, or hyperbolization can engender a distorted narrative that obfuscates regulatory oversight and misleads investors. AI-augmented ESG appraisal mechanisms may artificially embellish an entity's ecological performance by accentuating inconsequential eco-centric initiatives while clandestinely obfuscating large-scale environmental transgressions.[8]

b) **Algorithmic Bias in ESG Ratings**: Corporations increasingly deploy AI-infused ESG evaluative matrices to substantiate and propagate their ostensible sustainability credentials. However, the opacity inherent in algorithmic decision-making, compounded by the latent biases embedded within training corpora, frequently culminates in artificially inflated ESG indices. This engenders a spurious veneer of environmental probity that deceives both regulatory bodies and fiduciary stakeholders. A paradigmatic exemplar is Tesla's contentious exclusion from the S&P 500 ESG Index in 2022, despite its ostensible commitment to electric vehicular innovation, while petrochemical conglomerates, through strategic disclosures, maintained superior ESG standings.[9]

c) **Deepfake**: AI-powered marketing tools can fabricate hyper-realistic audiovisual content, falsely presenting a company as environmentally responsible. Such deceptive techniques exploit cognitive biases in consumer behaviour, making AI-driven greenwashing harder to detect.[10]

---

[1] Magali A. Delmas & Vanessa Cuerel Burbano, *The Drivers of Greenwashing*, 54 CAL. MGMT. REV. 64 (2011).

[2] Amanda Shanor & Sarah E. Light, *Greenwashing and the First Amendment*, 122 COLU. L. REV. 2033 (2022).

[3] Pham Ngoc Thinh, *Greenwashing and the Challenge of Sustainable Development in the Construction Industry*, 22 UD-JST 57 (2024).

[4] Nancy E. Furlow, *Greenwashing in the New Millennium*, 10 J. APPLIED BUS. & ECON. 22 (2010).

[5] Emmanuel Maalouf, *Achieving Corporate Environmental Responsibility through Emerging Sustainability Laws*, 27 ASIA PAC. J. ENVTL. L. 64 (2024).

[6] Felice Janice Olivia Boedijanto & Laurence L. Delina, *Potentials and Challenges of Artificial Intelligence-Supported Greenwashing Detection in the Energy Sector*, 115 ENERGY RES. & SOC. SCI. 103638 (2024).

---

[7] Honglei Mu & Youngchan Lee, *Greenwashing in Corporate Social Responsibility: A Dual-Faceted Analysis of Its Impact on Employee Trust and Identification*, 15 SUSTAINABILITY 15693 (2023).
[8] Radu Simion, *Eco-Frauds: The Ethics and Impact of Corporate Greenwashing*, 69 STUD. U. BABES-BOLYAI - PHIL. 7 (2024).
[9] RICHARD HARDYMENT, MEASURING GOOD BUSINESS: MAKING SENSE OF ENVIRONMENTAL, SOCIAL AND GOVERNANCE (ESG) DATA 150 (1 ed. 2024).
[10] Breda McCarthy, *Can Generative Artificial Intelligence Help or Hinder Sustainable Marketing?: An Overview of Its Applications, Limitations and Ethical Considerations*, 4 JRE 18 (2024).

**Why Does AI Obfuscate Criminal Culpability?**

AI-driven greenwashing engenders unprecedented jurisprudential quandaries within the domain of criminal liability.[11] The autonomous dynamism of AI decision-making, the absence of direct human volition, and the inherent complexities in ascribing culpability collectively confound traditional legal doctrines.[12] Unlike conventional greenwashing—where corporate executives or marketing consortia willfully orchestrate deceptive stratagems—AI-mediated greenwashing precipitates a confluence of legal conundrums[13]:

  i. **Absence of *Mens Rea* (Culpable Intent)**: The doctrinal bedrock of criminal liability necessitates the presence of *mens rea* (a culpable mental state), involving intent, cognisance, or recklessness. However, AI architectures operate without sentience, volitional agency, or subjective intentionality. Consequently, judicial scrutiny must ascertain whether culpability should be imputed to corporate entities, AI developers, or data curators.[14]

 ii. **Substantiating *Actus Reus* (Culpable Conduct)**: AI systems autonomously assimilate data, synthesise ESG disclosures, and propagate environmental assertions. This raises the pivotal interrogation of whether AI-generated misrepresentations constitute a deliberate malfeasance or a stochastic algorithmic aberration. The ontological distinction between intentional corporate malfeasance and mechanised computational fallibility further obfuscates the threshold for establishing criminal transgression.[15]

iii. **Corporate Criminal Liability**: Numerous legal regimes predicate corporate culpability upon the identification doctrine, wherein juridical responsibility is imputed to senior executives who embody the "directing mind and will" of the corporation. However, AI-induced greenwashing subverts this doctrinal construct. The algorithmic autonomy intrinsic to AI-generated decision-making often operates extrinsically to direct human intervention. This jurisprudential lacuna has impelled certain legal theorists to advocate for a novel doctrine of AI culpability.[16] Under this emergent doctrinal schema, corporate entities would incur strict liability for algorithmically orchestrated deceptions, irrespective of human intentionality or volitional misconduct.[17]

---

[11] Miriam Buiten, Alexandre De Streel & Martin Peitz, *The Law and Economics of AI Liability*, 48 COMP. L. & SEC. REV. 105794 (2023).

[12] Ramy El-Kady, *Challenges of Criminal Liability for Artificial Intelligence Systems:*, *in* ADVANCES IN ELECTRONIC GOVERNMENT, DIGITAL DIVIDE, AND REGIONAL DEVELOPMENT 1 (Halim Bajraktari ed., 2025).

[13] Rachelle Downie & Teisha Deckker, *Going Green: Green Trade Marks and the Risks of Greenwashing*, INTELLECTUAL PROPERTY FORUM: JOURNAL OF THE INTELLECTUAL AND INDUSTRIAL PROPERTY SOCIETY OF AUSTRALIA AND NEW ZEALAND 24 (2024).

[14] Mihailis E. Diamantis, *The Extended Corporate Mind: When Corporations Use AI to Break the Law*, 98 N.C. L. REV. 893 (2019).

[15] *Id*.

[16] CHRISTOPHER MARKOU & SIMON DEAKIN, IS LAW COMPUTABLE? 120 (2020).

iv. **Ethics**: Prevailing environmental statutes and anti-fraud regulations remain antiquated vis-à-vis AI-facilitated offences. The absence of meticulously delineated legal taxonomies governing AI culpability engenders a formidable enforcement deficit. This lacuna enables corporate entities to invoke algorithmic opacity as a juridical subterfuge, circumventing culpability by invoking plausible deniability concerning AI-generated misrepresentations. The resultant regulatory vacuum exacerbates AI ethics' opacity, allowing nefarious actors to exploit technological ambiguity as an exculpatory shield against legal reprisal.[18]

# II
## AI and the Machinations of Greenwashing

AI constitutes an instrumental force in the architectural formulation of corporate sustainability reports.[19] Many corporate entities leverage natural language processing (NLP) to synthesise reports ostensibly conforming to ecological benchmarks and regulatory edicts.[20] However, these algorithmically generated disclosures are often bereft of substantive accountability and epistemic transparency.[21] AI fabricates a simulacrum of corporate environmental responsibility by strategically accentuating favourable sustainability indices while obfuscating deleterious ecological externalities.[22] For instance, AI-driven models autonomously aggregate datasets from disparate sources to construct ESG reports highlighting compliance while omitting environmentally harmful practices. The absence of human oversight in these algorithmically engineered reports engenders profound epistemological and regulatory trepidations. The veracity of AI-curated sustainability narratives remains inherently dubious, impeding regulators and fiduciary stakeholders from delineating authentic ecological commitments from algorithmically orchestrated subterfuge.[23] ESG ratings, which assess a company's environmental and social impact, are increasingly influenced by AI algorithms. AI-driven ESG analytics aggregate vast amounts of corporate data to generate ratings that impact investor decisions. However, these systems are susceptible to manipulation. Corporations can strategically feed AI-selected datasets that inflate their ESG performance while omitting detrimental practices.[24] A notable example is the discrepancy in ESG ratings across different providers. Research shows that AI-generated ESG scores can vary widely depending on the data inputs and weighting methods used by rating agencies.[25] This inconsistency raises concerns about the reliability of AI-driven ESG assessments, as firms can exploit algorithmic opacity to mislead. Beyond ESG manipulation, AI is used to misrepresent corporate environmental practices through automated digital marketing, chatbot responses, and algorithmically curated content. AI-driven sentiment analysis tools scan media and online discussions to shape corporate narratives. This often suppresses negative publicity, amplifying sustainability claims. Social media platforms further exacerbate AI-driven greenwashing by using algorithmic promotion to prioritise corporate sustainability messages while deprioritising reports on environmental violations. This creates an informational asymmetry where consumers and investors encounter curated, AI-optimised green narratives rather than objective ecological performance data.[26]

---

[17] Rossella Sabia, *Artificial Intelligence and Environmental Criminal Compliance*, in THE CRIMINAL LAW PROTECTION OF OUR COMMON HOME 179 (2019).

[18] Simion, *supra* note 10.

[19] Wayne Moodaley & Arnesh Telukdarie, *Greenwashing, Sustainability Reporting, and Artificial Intelligence: A Systematic Literature Review*, 15 SUSTAINABILITY 1481 (2023).

[20] Nils Smeuninx, Bernard De Clerck & Walter Aerts, *Measuring the Readability of Sustainability Reports: A Corpus-Based Analysis Through Standard Formulae and NLP*, 57 INT'L J. BUS. COMM. 52 (2020).

[21] Ellen Pei-yi Yu, Bac Van Luu & Catherine Huirong Chen, *Greenwashing in Environmental, Social and Governance Disclosures*, 52 RES. INT'L BUS. & FIN. 101192 (2020).

[22] Roberto Rodrigues Loiola & Ludmila de Melo Souza, *Greenwashing and Corporate Sustainability: A Systematic Literature Review Focusing on AI and Machine Learning Applications*, SEMEAD (2024).

Although not an archetypal AI-induced malfeasance, the Volkswagen emissions debacle (*Dieselgate*) constitutes a seminal jurisprudential precedent elucidating the pernicious ramifications of technological subterfuge in ecological deception. Volkswagen surreptitiously embedded software within its diesel-powered vehicles, thereby effectuating the falsification of emissions diagnostics and engendering a grand-scale misrepresentation of environmental compliance. This stratagem precipitated the systemic obfuscation of regulatory scrutiny and the calculated misdirection of consumer perception regarding the ecological impact of its vehicles.[27] This case epitomises the latent capacity of automated computational frameworks to perpetuate industrial-scale greenwashing. Had AI-driven mechanisms been deployed to fabricate emissions analytics or algorithmically automate regulatory compliance documentation, the magnitude of deception could have been exponentially exacerbated. AI-augmented reporting possesses the potential to engender a hyper-sophisticated façade of sustainability, thereby rendering ecological malfeasance impervious to conventional oversight mechanisms. This precedent underscores the need for stringent oversight over AI-mediated corporate disclosures to forestall algorithmic manipulation of sustainability taxonomies.

---

[23] Nichole Li et al., *Using Artificial Intelligence in ESG Assurance*, 21 J EMERGING TECH. ACCT. 83 (2024).

[24] Martina Macpherson, Andrea Gasperini & Matteo Bosco, *Implications for Artificial Intelligence and ESG Data*, (2021).

[25] Florian Berg, Julian F. Kölbel & Roberto Rigobon, *Aggregate Confusion: The Divergence of ESG Ratings*, (2019).

[26] Rongxin Chen & Tianxing Zhang, *Artificial Intelligence Applications Implication for ESG Performance: Can Digital Transformation of Enterprises Promote Sustainable Development?*, CMS (2024).

[27] *United States v. Volkswagen AG*, No. 16-CR-20394 (E.D. Mich. 2016).

## III
## Criminal Law Governing AI-Induced Greenwashing

AI-mediated greenwashing engenders profound juridical inquiries about fraud and material misrepresentation, as corporate entities increasingly harness artificial intelligence to hyperbolise ecological bona fides.[28] Fraud, canonically delineated as an intentional act of deception designed to procure illicit or inequitable advantage, may encapsulate AI-orchestrated fabrications of sustainability credentials.[29] The juridical applicability of fraud laws to AI-generated ecological disinformation is contingent upon attributing culpability to human agents within corporate hierarchies. If AI autonomously engender deceptive sustainability proclamations, the ascertainment of *mens rea* (criminal intent) becomes an intricate doctrinal conundrum. The absence of sentient volition in algorithmic constructs further complicates the imposition of conventional criminal liability.[30] Several existing laws provide avenues for addressing such deception:

i. **United Kingdom**: The Fraud Act 2006 statutorily proscribes the promulgation of spurious representations effectuated with the intent to procure an unjust pecuniary advantage or precipitate financial detriment. The Act's Sec. 2 could apply to corporations using AI to disseminate misleading environmental claims.[31]

ii. **US:** The Securities Exchange Act of 1934, in conjunction with the regulatory edicts promulgated by the SEC[32], categorically interdicts fraudulent disclosures within corporate expositions. Additionally, the FTC[33] Green Guides furnish a supplementary enforcement apparatus designed to counteract spurious environmental proclamations.[34]

---

[28] Peter Dauvergne, *Is Artificial Intelligence Greening Global Supply Chains? Exposing the Political Economy of Environmental Costs*, 29 REV. INT'L POL. ECON. 696 (2022).

[29] Muhammad Kaleem Khan et al., *The Automated Sustainability Auditor: Does Artificial Intelligence Curtail Greenwashing Behavior in Chinese Firms?*, 33 BUS STRAT ENV 9015 (2024).

[30] Mohamed Fathi Shehta Diab, *Criminal Liability for Artificial Intelligence and Autonomous Systems*, 3 AM. J. SOC. L. 14 (2024).

[31] Adam Bernstein, *Putting Fraudsters on Notice*, 12 J. OF AESTHETIC NURSING 422 (2023).

[32] Securities and Exchange Commission.

[33] Federal Trade Commission.

   iii. **India:** The Indian Penal Code, 1860, criminalises fraudulent inducement and deceptive misrepresentation under Sec. 420, which proscribes acts of cheating effected through dishonestly induced transactions. Moreover, the Consumer Protection Act, 2019, vests the CCPA[35] with expansive adjudicatory prerogatives to proscribe deceptive advertising, a provision that could potentially subsume AI-mediated greenwashing within its regulatory purview. [36]

Precedents Illustrating the Enforcement of Anti-Fraud Statutes in Environmental Misrepresentation:

   i. **Federal Trade Commission v. Volkswagen AG**: This critical case delineated corporate culpability for orchestrating fraudulent vehicular emissions disclosures. Although not predicated upon AI-driven subterfuge, the case underscored the judiciary's propensity to impose liability for technologically manipulated environmental misrepresentations.[37]

   ii. **Securities and Exchange Commission v. Vale S.A.**: This litigation encapsulated the regulatory censure of fraudulent ESG disclosures, epitomising the intensification of oversight mechanisms governing corporate greenwashing. The ruling reaffirmed the imperative of stringent compliance protocols in corporate sustainability representations, reinforcing the SEC's prerogative to adjudicate deceptive environmental proclamations.[38]

Since AI lacks legal personhood, corporate criminal liability doctrines must be adapted to address AI-driven greenwashing. The identification doctrine, which holds a corporation liable by attributing criminal conduct to its senior managers, presents limitations in AI-related cases. Courts may struggle to identify a natural person responsible for an AI-generated misrepresentation, particularly when decision-making processes involve machine learning algorithms without direct human input.[39]

---

[34] Florian Berg, Julian F Kölbel & Roberto Rigobon, *Aggregate Confusion: The Divergence of ESG Ratings*, 26 REVIEW OF FINANCE 1315 (2022).
[35] Central Consumer Protection Authority.
[36] Dinesh Kumar, *Tackling Greenwashing in Global Environment, Social and Governance Reporting: A Legal and Technical Perspective on Corporate Accountability*, (2024).
[37] *FTC v. Volkswagen AG*, No. 16-cv-01534, 2016 WL 4376623 (N.D. Cal. 2016).
[38] *SEC v. Vale S.A.*, No. 22-cv-2407, 2022 WL 16837080 (E.D.N.Y. 2022).
[39] Deny Setiawan, Warasman Marbun & Arief Patramijaya, *Corporate Criminal Liability In Environmental Pollution Crimes*, 5 JIRPL 511 (2024).

i. **United Kingdom:** As enunciated in Tesco Supermarkets Ltd v. Nattrass, the identification doctrine within English jurisprudence circumscribes corporate culpability to senior functionaries who constitute the entity's "directing mind and will".[40] This doctrinal construct predicates liability upon human agency, thereby rendering AI-induced infractions a complex adjudicatory conundrum.

   ii. **United States:** The respondeat superior doctrine imputes corporate liability for the malfeasance of employees, provided such infractions transpire within the ambit of occupational functions.[41] This expansive liability paradigm posits that corporate entities may bear vicarious culpability for AI-generated greenwashing, contingent upon judicial recognition of AI as an operational extension of corporate instrumentalities.

   iii. **India:** Vicarious liability enshrined in the Companies Act mandates corporate culpability for fraudulent contraventions effected by corporate officers.[42] This statutory edifice predicates liability on hierarchical accountability, thereby necessitating doctrinal recalibration to accommodate AI-autonomous deception.

Given AI's algorithmic autonomy, courts may invoke strict liability precepts to ascribe culpability to corporate entities for AI-mediated greenwashing, irrespective of volitional intent. The EU CSDDD [43] envisages an augmented due diligence apparatus governing sustainability expositions.[44] AI-generated greenwashing may be subsumed under environmental crime statutes, should algorithmically contrived misrepresentations precipitate tangible ecological degradation.[45] Environmental crime statutes conventionally predicate liability upon direct ecological degradation or illicit pollutive infractions. However, the juridical ambit of such laws may extend to encompass deceptive sustainability representations under expansive environmental misrepresentation provisions.

i. **United States**: The Clean Air Act [46] and the FTC Act [47] proscribe corporate malfeasance predicated upon fraudulent ecological assertions. These statutory instruments empower regulatory agencies to censure and penalise corporate entities that promulgate fallacious green credentials, thereby subverting market integrity and consumer trust.

---

[40] *Tesco Supermarkets Ltd. v. Nattrass*, [1972] A.C. 153 (H.L.).
[41] *New York Central & Hudson River Railroad Co. v. United States*, 212 U.S. 481 (1909).
[42] Companies Act, No. 18 of 2013, § 447 (India).
[43] Corporate Sustainability Due Diligence Directive.
[44] Proposal for a Directive of the European Parliament and of the Council on Corporate Sustainability Due Diligence, COM (2022) 71 final (Feb. 23, 2022).
[45] Mitzi Bolton & Tom Chan, *Governing Sustainability Transitions in a World of AI-Powered Greenwashing*, (2024).
[46] Clean Air Act, 42 U.S.C. § 7401 (2018).
[47] Federal Trade Commission Act, 15 U.S.C. §§ 41–58 (2018).

ii. **European Union**: The EU Green Claims Directive (Proposed) envisages a regulatory apparatus meticulously calibrated to obviate misleading sustainability proclamations. This legislative initiative seeks to impose enhanced due diligence obligations upon corporate disclosures, ensuring verifiable and substantiated environmental assertions.[48]

iii. **India**: The Environmental Protection Act, under Sec. 15 and 16, prescribes stringent penal sanctions for ecological misrepresentation. This statutory architecture establishes punitive consequences for deceptive environmental declarations that precipitate regulatory subversion or ecological harm.[49]

While existing criminal laws provide partial remedies for AI-driven greenwashing, the attribution of liability remains a challenge. Fraud and corporate criminal liability doctrines offer prosecutorial avenues, but AI's autonomous decision-making complicates *mens rea* assessment. Environmental crime laws may supplement enforcement efforts, particularly where greenwashing results in tangible ecological harm. Future legal reforms must ensure that corporations deploying AI for sustainability claims remain accountable under criminal law.

## IV
## Challenges In Attributing Criminal Liability

The juridical application of criminal liability to AI-mediated greenwashing engenders doctrinal and pragmatic complexities. The foundational precepts of criminal culpability—*mens rea* (culpable mental state) and *actus reus* (proscribed conduct)—become doctrinally convoluted when adjudicating the malfeasance of artificial intelligence.[50] AI operates devoid of sentient cognition, volitional intentionality, or conventional human agency, thereby subverting traditional paradigms of penal attribution. The ontological absence of subjective criminal intent within algorithmic constructs necessitates a jurisprudential recalibration to accommodate the algorithmic orchestration of environmental misrepresentation. This section examines these complexities, evaluates legal frameworks, and explores whether AI could be recognised as a legal person.[51]

---

[48] Proposal for a Directive of the European Parliament and of the Council on Substantiation and Communication of Explicit Environmental Claims (Green Claims Directive), COM (2023) 166 final (Mar. 22, 2023).

[49] The Environment (Protection) Act, No. 29 of 1986, §§ 15–16 (India).

[50] Amy McGovern et al., *Why We Need to Focus on Developing Ethical, Responsible, and Trustworthy Artificial Intelligence Approaches for Environmental Science*, 1 ENVIRON. DATA SCIENCE e6 (2022).

### A. *Mens Rea* (Intent) in AI-Driven Crime

*The Absence of Human Intent in AI Actions*

*Mens rea*, the sine qua non of criminal culpability, constitutes an indispensable doctrinal pillar within the edifice of penal jurisprudence. Canonical fraud statutes necessitate demonstrably substantiating fraudulent animus, mandating evidentiary corroboration of deliberate and premeditated intent to deceive. AI-driven greenwashing, however, complicates this requirement, as AI lacks subjective intent.[52]

a) AI functions through preordained algorithmic architectures, autonomously evolving machine learning, and data-driven computational heuristics, thereby obfuscating the attribution of conventional *mens rea* to any discrete juridical or corporate entity.[53]

b) When an AI-driven apparatus fabricates spurious sustainability expositions or artificially inflated ESG indices, the juridical delineation of its conduct as intentional malfeasance, reckless endangerment, or culpable negligence becomes an intricate adjudicatory enigma, ensnared in layers of doctrinal ambiguity and evidentiary opacity.[54]

*Corporate Mens Rea and Attribution of AI Intent*

If an AI-driven system generates fraudulent environmental claims, criminal liability may shift to the corporation deploying it. The Identification Doctrine—which holds corporate officers personally liable for corporate crimes—faces limitations: the UK House of Lords restricted corporate liability to individuals who constitute the "directing mind and will" of the company.[55] If an AI system independently generates misleading sustainability claims, the absence of a culpable corporate officer could hinder prosecution. Some jurisdictions employ strict liability to bypass the *mens rea* requirement. Under statutes like the US Green Guides (16 CFR Part 260), regulators hold corporations accountable for misleading environmental claims regardless of

---

[51] Alaa Saud, *Criminal Liability about the Use of Artificial Intelligence: Investigating the Actus Reus Element of AI-Driven Technology*, 6 AJL 1 (2024).

[52] Monika Simmler & Nora Markwalder, *Guilty Robots? – Rethinking the Nature of Culpability and Legal Personhood in an Age of Artificial Intelligence*, 30 CRIM LAW FORUM 1 (2019).

[53] Sylwia Wojtczak, *Endowing Artificial Intelligence with Legal Subjectivity*, 37 AI & SOC 205 (2022).

[54] Zhuozhen Duan, *Artificial Intelligence and the Law: Cybercrime and Criminal Liability*, 62 THE BRITISH JOURNAL OF CRIMINOLOGY 257 (2022).

[55] *Tesco Supermarkets Ltd. v. Nattrass*, [1972] A.C. 153 (H.L.).

intent. However, criminal law traditionally resists strict liability for fraud-based offences.[56]

*Recklessness and Negligence in AI-Driven Fraud*

In the juridical void left by AI's incapacity for *mens rea*, prosecutorial arguments may pivot towards recklessness or negligence as alternative culpability:

i. **Recklessness**: If a corporate entity deliberately operationalises an AI system with a demonstrably elevated propensity for engendering deceptive environmental analytics, such conduct may satisfy the recklessness threshold enshrined within statutory fraud architectures. The foreseeability of algorithmic malfeasance and conscious disregard for its deleterious ramifications constitute a pivotal axis of legal culpability.[57]

ii. **Negligence**: A corporate entity's dereliction of its fiduciary and regulatory obligations in supervising AI-generated sustainability expositions may precipitate civil or criminal liability under negligence-based doctrines. The failure to exercise requisite due diligence and algorithmic oversight in mitigating AI-induced environmental misrepresentation could invoke punitive repercussions.[58]

The House of Lords rearticulated the doctrinal contours of recklessness within criminal jurisprudence, accentuating the necessity of subjective risk cognisance.[59] If AI-facilitated ecological deception attains predictability through emergent regulatory scrutiny, corporate entities may incur liability for reckless greenwashing, predicated upon their conscious acquiescence to algorithmic malfeasance.

### B. *Actus Reus* (Guilty Act) in AI Greenwashing

*AI-Generated Deception as a Criminal Act*

The *actus reus* underpinning greenwashing materialises through the promulgation of spurious environmental assertions that engender regulatory obfuscation and mislead investors, consumers, or oversight bodies**.** The key questions are:

---

[56] Philipp Krueger et al., *The Effects of Mandatory ESG Disclosure Around the World*, 62 J OF ACCOUNTING RESEARCH 1795 (2024).

[57] Mario D. Schultz, Ludovico Giacomo Conti & Peter Seele, *Digital Ethicswashing: A Systematic Review and a Process-Perception-Outcome Framework*, AI ETHICS (2024).

[58] Vita Mahardhika, Pudji Astuti & Aminuddin Mustaffa, *Could Artificial Intelligence Be the Subject of Criminal Law?*, 12 YST 1 (2023).

[59] *R v. G*, [2003] UKHL 50, [2004] 1 A.C. 1034 (H.L.).

> Is AI-generated misinformation an "act" in criminal law?
> If AI autonomously produces misleading ESG reports, can that be classified as criminal?

In *United States v. Weitzenhoff*, judicial interpretation affirmed criminal liability for environmental infractions absent explicit *mens rea*, underscoring the doctrinal primacy of strict regulatory enforcement. The adjudicatory rationale employed therein could furnish a jurisprudential predicate for extending analogous liability principles to AI-mediated greenwashing, wherein corporate entities may incur culpability for algorithmic deception, irrespective of volitional intent.[60]

*Corporate Liability for AI-Generated Misrepresentation*

Corporate liability frameworks exhibit jurisdictional heterogeneity, with varying statutory architectures delineating the contours of fraudulent misrepresentation and regulatory contraventions:

i. **UK**: The Fraud Act, specifically Sec 2 (Fraud by False Representation), statutorily criminalises the willful propagation of mendacious declarations effectuated to procure pecuniary or strategic advantage.[61] In the context of AI-generated environmental disclosures, the principal juridical impediment lies in substantiating "knowledge" or "dishonesty" attributable to corporate functionaries, given the algorithmic autonomy intrinsic to AI-driven reportage.

ii. **United States**: Securities fraud statutes, notably Rule 10b-5 of the Securities Exchange Act of 1934[62], proscribe fraudulent machinations engineered to distort financial market equilibrium.[63] The algorithmic orchestration of ESG inflation—where AI-fabricated sustainability indices induce investor reliance on deceptive eco-centric proclamations—may precipitate liability under securities fraud doctrines.[64]

iii. **India**: The Environmental Protection Act[65], in conjunction with Sec 420 of IPC (Cheating and Dishonest Inducement)[66], furnishes a statutory mechanism for adjudicating deceptive sustainability representations. The confluence of environmental regulatory mandates and penal fraud statutes enables the criminalisation of corporate ecological misrepresentation, contingent upon evidentiary substantiation of culpable intent or reckless disregard for veracity.[67]

---

[60] *United States v. Weitzenhoff*, 35 F.3d 1275 (9th Cir. 1994).
[61] Fraud Act 2006, c. 35, § 2 (UK).
[62] 17 C.F.R. § 240.10b-5 (2023).
[63] Securities Exchange Act of 1934, 15 U.S.C. § 78j(b) (2018).
[64] Vicențiu-Traian Râmniceanu, *5. Vicentiu Ramniceanu*, 2 INT'L INV. L. J. 83 (2022).
[65] The Environment (Protection) Act, No. 29 of 1986, § 15 (India).
[66] Indian Penal Code, No. 45 of 1860, § 420 (India).
[67] Kazimieras Simonavičius University, Lithuania & Giedrius Nemeikšis, *Artificial Intelligence As Legal Entity In The Civil Liability Context*, 12 AP 89 (2021).

### C. Legal Personhood of AI: Can AI Be Held Liable?

*The Debate Over AI Legal Personality*

Certain legal theorists advocate for the conferral of juridical personhood upon AI systems, akin to the doctrinal construct under which corporate entities are vested with distinct legal personhood. This framework would ostensibly facilitate the litigation or penal sanctioning of AI-driven malfeasance as an autonomous juridical entity. The European Parliament's Resolution on Civil Law Rules on Robotics (2017/2103(INL)) posited the conceptualisation of "electronic personhood" as a prospective legal taxonomy for autonomous AI systems. This legislative discourse envisaged an independent legal identity for AI, thereby enabling its subjection to regulatory and penal accountability.[68] Opponents of this proposition contend that such a doctrinal realignment is jurisprudentially untenable and pragmatically deficient. AI lacks sentient cognition or proprietary assets and lacks the corporeal agency requisite for conventional penal repercussions. The incapacity of AI to be incarcerated or financially sanctioned independent of its corporate proprietors underscores the enforcement deficit inherent in this conceptual framework.

*AI as a Tool vs. AI as an Offender*

An alternative doctrinal framework conceptualises AI as a mechanised adjunct to corporate operations rather than a distinct juridical malefactor. Legal culpability is imputed to corporate entities or AI developers, who exercise dominion over algorithmic functionalities and operational deployment. The Supreme Court [69] reaffirmed the responsible corporate officer doctrine, holding senior executives vicariously culpable for regulatory infractions under their administrative purview. This doctrinal formulation could be extrapolated to AI-mediated corporate transgressions, wherein executives overseeing AI-driven compliance mechanisms may incur derivative liability for algorithmic malfeasance.[70] While Indian jurisprudence has yet to adjudicate AI culpability explicitly, extant corporate liability doctrines under the Companies Act[71] furnish a potential adjudicatory basis for AI-induced infractions.[72] In principle, the statutory imposition of vicarious liability upon corporate hierarchies could be jurisprudentially extended to encompass AI-facilitated regulatory contraventions. The juridical attribution of criminal liability for AI-mediated greenwashing is encumbered by formidable doctrinal impediments. The absence of human volitional intent, the evidentiary conundrum of establishing corporate *mens rea*, and the escalating entrenchment of AI within corporate decision-making architectures collectively engender adjudicatory ambiguity. Nevertheless, emergent legal paradigms proffer prospective resolutions:

1) **Corporate Accountability Constructs**: The fortification of corporate liability statutes to explicitly encompass AI-induced fraudulent misrepresentation and algorithmic deception within their penal ambit.[73]

2) **Mandated AI Due Diligence**: The imposition of statutorily enshrined regulatory imperatives obligating continuous oversight, audit mechanisms, and compliance verification for AI-driven reporting under ESG and fraud statutes.[74]

3) **Expansive Strict Liability Doctrines**: The jurisprudential extension of strict liability frameworks to encapsulate AI-generated ecological falsifications, ensuring culpability irrespective of intent or knowledge.[75]

4) **Judicial Doctrinal Evolution**: The reconfiguration of established legal doctrines, encompassing intent, recklessness, and vicarious liability, to accommodate the technological singularities of AI-mediated corporate malfeasance.

---

[68] CIVIL LAW RULES ON ROBOTICS (2015/2103(INL)), 23 (2019).

[69] *United States v. Park*, 421 U.S. 658 (1975).

[70] Ross Bellaby, *The Ethical Problems of 'Intelligence–AI,'* 100 INT'L AFFAIRS 2525 (2024).

[71] Companies Act, No. 18 of 2013, § 447 (India).

[72] Fabio Caputo et al., *Enhancing Environmental Information Transparency through Corporate Social Responsibility Reporting Regulation*, 30 BUS STRAT ENV 3470 (2021).

As AI-driven automation further metastasises within corporate governance ecosystems, global legal frameworks must undergo proactive recalibration to forestall the circumvention of regulatory oversight. The convergence of AI, corporate fraud, and environmental jurisprudence necessitates a robust, adaptive, and doctrinally fortified legal infrastructure to impute liability and safeguard regulatory integrity.

# V

# Conclusion

AI-driven greenwashing epitomises an emergent jurisprudential conundrum that subverts traditional liability doctrines, engendering regulatory lacunae in environmental fraud adjudication. The anthropocentric predicates of existing criminal laws, predicated on demonstrable *mens rea* and hierarchical corporate culpability, are rendered ontologically deficient when confronted with algorithmic opacity and autonomous deception. This study's comparative dissection of statutory architectures across India, the US, and the EU underscores the doctrinal obsolescence of extant fraud statutes in encapsulating AI-induced misrepresentation. Preliminary findings substantiate the exigency for a hybrid liability construct integrating strict liability precepts, algorithmic due diligence imperatives, and recalibrated corporate governance taxonomies to circumvent prosecutorial impediments posed by AI's volitional vacuum. The EU Corporate Sustainability Due Diligence Directive emerges as a paradigmatic precedent for transnational regulatory harmonisation, offering a potential juridical scaffolding to preempt algorithmic subterfuge in sustainability disclosures. This study advances the intellectual corpus of AI governance and environmental jurisprudence by advocating for a doctrinal evolution that transcends anachronistic liability paradigms, ensuring that AI-driven ecological deception is neither extrajudicially insulated nor juridically indeterminate.

---

[73] Judit Bayer, *Legal Implications of Using Generative AI in the Media*, 33 INFO. & COMM. TECH. L. 310 (2024).

[74] Serena Oduro, Emanuel Moss & Jacob Metcalf, *Obligations to Assess: Recent Trends in AI Accountability Regulations*, 3 PATTERNS 100608 (2022).

[75] Rushil Chandra & Karun Sanjaya, *Punishing the Unpunishable: A Liability Framework for Artificial Intelligence Systems*, *in* DIGITAL TECHNOLOGIES AND APPLICATIONS 55 (Saad Motahhir & Badre Bossoufi eds., 2023).

The entrenchment of AI within corporate decision-making architectures engenders jurisprudential conundrums for traditional legal taxonomies, particularly in criminal culpability vis-à-vis greenwashing. Across multiple jurisdictions, the cardinal impediment in adjudicating AI-induced corporate fraud is the ascription of liability, especially in scenarios where direct human volition is absent in fabricating spurious or deceptive environmental representations. Extant legal infrastructures, including fraud statutes and environmental regulatory mandates, were not architected to contend with autonomous algorithmic mechanisms capable of orchestrating financial misrepresentation, distorting ESG indices, and propagating contrived sustainability narratives, all bereft of direct human intervention. This juridical lacuna necessitates an expansive doctrinal recalibration, ensuring that AI-driven ecological deception does not subvert regulatory oversight or evade prosecutorial scrutiny. To ameliorate the doctrinal and regulatory lacunae precipitated by AI-driven ecological misrepresentation, legal frameworks must undergo meticulous recalibration to accommodate the technological singularities of algorithmic deception. The following juridical and legislative modifications warrant consideration:

a. **Conferral of Juridical Personhood Upon AI for Liability Attribution**: As AI systems progressively attain operational autonomy, a pivotal reformative trajectory could entail the juridical recognition of AI as a distinct legal entity, thereby rendering it susceptible to liability adjudication in environmental fraud. Contemporary corporate liability constructs predicate culpability upon natural persons or juridical corporate entities, a paradigm that remains jurisprudentially inadequate when AI operates as the primary architect of deception. The institutionalisation of electronic personhood would necessitate a fundamental reconfiguration of existing legal definitions, corporate accountability doctrines, and adjudicatory mechanisms.

b. **Legislative Augmentation of Fraud and Environmental Statutes**: Existing fraud statutes, including the UK Fraud Act 2006, the U.S. Securities Exchange

Act, and India's Penal Code provisions on fraudulent misrepresentation, are anthropocentric in orientation, necessitating reformulation to encapsulate AI-orchestrated deception. A legislative recalibration could empower regulatory authorities, such as the U.S. SEC, to explicitly subsume AI-facilitated greenwashing within anti-fraud enforcement mechanisms. Analogously, the UK Fraud Act could be retrofitted to accommodate AI-generated misinformation concerning corporate sustainability claims, thereby expanding the definitional breadth of fraud within environmental misrepresentation**.**

c. **Institutionalisation of Strict Liability Doctrines for AI-Induced Deception**: The imposition of strict liability on corporate entities leveraging AI systems could circumvent evidentiary impediments associated with establishing intent-based culpability in AI-generated fraud. This doctrinal evolution would ensure that corporate entities remain juridically accountable for algorithmic misrepresentation, irrespective of human intent. The absence of a volitional actor in AI deception should not serve as an exculpatory shield against regulatory intervention. Laws would compel corporations to institute robust algorithmic oversight mechanisms to mitigate fraudulent environmental claims by mandating strict liability.

d. **Reconfiguration of Corporate Governance**: A paradigmatic overhaul of corporate governance infrastructures is requisite to institutionalise AI ethics, algorithmic transparency, and sustainability-centric oversight. This would necessitate the formulation of compliance architectures encompassing AI-centric audit mechanisms, algorithmic transparency disclosures, and third-party regulatory scrutiny. Furthermore, codifying AI governance prerogatives within corporate leadership echelons—such as establishing board-level oversight committees specialising in AI-induced environmental liabilities—would ensure that corporate decision-making adheres to ethical and regulatory sustainability imperatives.

By synthesising these, laws can preemptively fortify accountability mechanisms against AI-driven environmental deception, thereby foreclosing regulatory lacunae that would otherwise enable algorithmic subterfuge to evade juridical scrutiny. The exigency of a comprehensive AI liability architecture is paramount to ameliorate the doctrinal voids engendered by AI-driven ecological deception. Such a regulatory edifice must reconcile corporate culpability imperatives with the ontological complexities of algorithmic autonomy. Given the intrinsically transnational character of corporate operations and environmental exigencies, multilateral legal harmonisation is indispensable in mitigating AI-mediated environmental fraud and fortifying juridical accountability mechanisms. To preempt regulatory fragmentation, sovereign entities must synchronise legal taxonomies through global juridical instruments, codifying uniform enforcement paradigms tailored to AI's role in sustainability misrepresentation. This necessitates the institutionalisation of international treaties or multilateral legal accords delineating cohesive AI governance and liability matrices.

An archetypal precedent of juridical transnationalism is the EU CSDDD, which mandates corporate due diligence in environmental, human rights, and governance spectrums. This legislative corpus could be a paradigmatic template for global regulatory architectures, integrating AI accountability within corporate governance taxonomies and environmental jurisprudence. Moreover, the UNEP [76] and the OECD[77], could operationalise globally cohesive regulatory guidelines to counteract AI-induced greenwashing and corporate ecological fraud. Additionally, experts could synthesise a comprehensive doctrinal schema to entrench algorithmic transparency and liability principles into global environmental governance. This would preemptively eliminate regulatory vacuums and ensure that corporate entities remain juridically accountable, irrespective of territorial jurisdiction.

---

[76] United Nations Environment Programme.
[77] Organisation for Economic Co-operation and Development.